\documentclass[twocolumn]{jpsj3}
\usepackage{txfonts}

\title{Intrinsic Proton NMR Studies of Mg(OH)$_2$ and Ca(OH)$_2$}

\author{Yutaka Itoh$^1$\thanks{E-mail:yitoh@cc.kyoto-su.ac.jp} and Masahiko Isobe$^2$}
\inst{$^1$Department of Physics, Graduate School of Science, Kyoto Sangyo University, Kyoto 603-8555, Japan \\
$^2$Max-Planck-Institut f\"{u}r Festk\"{o}rperforschung, Heisenbergstrasse 1, D-70569 Stuttgart, Germany } %\\

\abst{
We studied the short proton free induction decay signals and the broad $^1$H NMR spectra of Mg(OH)$_2$ and Ca(OH)$_2$ powders at 77$-$355 K and 42 MHz
using pulsed NMR techniques. 
Using a Gaussian-type back extrapolation procedure for the obscured data of the proton free induction decay signals,
we obtained more precise values of the second moments of the Fourier-transformed broad NMR spectra than those in a previous report [Y. Itoh and M. Isobe, J. Phys. Soc. Jpn. {\bf 84}, 113601 (2015)] and compared with the theoretical second moments. 
The decrease in the second moment could not account for the large decrease in the magnitude of the intrinsic proton spin-lattice relaxation rate 1/$T_1$ from Mg(OH)$_2$ to Ca(OH)$_2$. 
The analysis of 1/$T_1$ $\propto$ exp($-E_g$/$k_\mathrm{B}T$) with $E_g$ $\sim$ 0.01 eV points to a local hopping mechanism, and
that of 1/$T_1 \propto T^{n}$ with $n\sim$ 0.5 points to an anharmonic rattling mechanism.
}
 
\begin{document}
\maketitle
  
\section{Introduction}  
The divalent hydroxides of M(OH)$_2$ (M = Mg, Ca) have a CdI$_2$-type structure with broken site-symmetry at the hydrogen site.
The atomic displacement parameters of the proton at the 6$i$ Wyckoff site ($x$, 2$x$, $z$) in $P\overline{3}$
are large~\cite{ND2,NDCa} and comparable to those of the rattling ions in the cage structures of pyrochlore systems~\cite{ZHiroi}. 
Theoretical computer analyses indicate anharmonic motion and temperature-dependent angular correlation of the OH groups~\cite{DFT,MD}.  

In recent pulsed NMR experiments for Mg(OH)$_2$ and Ca(OH)$_2$ powders~\cite{Itoh},
we have found that the free induction decay (FID) of the proton magnetization is a superposition of a short decay in $\sim$ 30 $\mu$s and a long decay in $\sim$ 1 ms.  
Then, the corresponding frequency spectra show the superposition of broad and narrow components,
which are assigned to intrinsic immobile protons and extrinsic mobile protons, respectively~\cite{Itoh,Itoh2}. 
 
Long before our reports, the pulsed NMR studies on proton motion showed nonexponential proton spin-lattice relaxation curves~\cite{T1Mg,T1Ca} and a slowing down effect for Ca(OH)$_2$~\cite{T1Ca}. 
However, our pulsed NMR studies revealed the nonexponential relaxation of the long FIDs (narrow components in the frequency spectra)
and the single/nearly single exponential relaxations of the short FIDs (broad components in the frequency spectra) for Mg(OH)$_2$ and Ca(OH)$_2$~\cite{Itoh,Itoh2}.
No slowing down effects were reproduced.  
The details of our NMR studies on the narrow components were reported in Ref. 7. 
Now, we revisit the broad components and the NMR analysis for the intrinsic proton NMR signals.   

The proton dipole field is so strong causing a short FID of the proton magnetization (nuclear spin $I$ = 1/2) in hydrogen compounds. 
In the actual pulsed NMR experiment, 
an obscure portion of the FID following an excitation $rf$ pulse prevents us from obtaining the full lineshape of a Fourier-transformed (FT) frequency spectrum
because of the lack of the short-time decay.  
How to handle the obscured portion of the FID following the excitation pulse is an issue in Fourier analysis~\cite{GBNMR}. 
The time shifting of a single exponential FID placed at zero time leads to a Lorentzian spectrum regardless of the zero.  
However, a time-shifted Gaussian decay does not lead to a simple Gaussian lineshape.
The dead time data of the FID also prevents us from obtaining the precise value of the second moment of the FT NMR spectrum.
 
\begin{figure}[b]
 \begin{center}
 \includegraphics[width=0.85\linewidth]{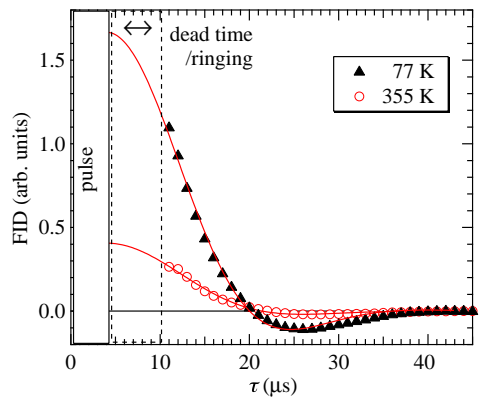}
 \end{center}
 \caption{\label{f1}
(Color online) Proton free induction decay (FID) signals $F(\tau)$ for Mg(OH)$_2$ powder at 77  and 355 K.  
The length of an excitation $rf$ pulse is 4 $\mu$s. The FID below 10 $\mu$s is obscured.  
Solid curves are the results from the least-squares fitting using an analytical function given by Eq.~(\ref{eq1}).
}
 \end{figure}
 
In Mg(OH)$_2$ and Ca(OH)$_2$ powders,
the short proton magnetization decays with slight oscillation~\cite{Itoh}.  
Figure~\ref{f1} shows the proton FID signals $F(\tau)$ for the Mg(OH)$_2$ powder at 77  and 355 K, 
where $\tau$ is the time following an excitation $rf$ pulse,
at a Larmor frequency $\nu_\mathrm{L}$ of 42.5772 MHz in an external magnetic field of 1.0 T.
The sample preparation and characterization have been carried out in previous studies~\cite{Itoh,Itoh2}.  
The oscillation is a Lowe beat in the FID due to a nuclear dipole$-$dipole interaction\cite{Lowe}.
The oscillating decay can be traced back to the magnetic dipolar lineshape of a more rectangular type than a Gaussian type~\cite{Abragam}. 
The short part of the FID ($\tau$ $<$ 10 $\mu$s) is obscured by probe ringing and/or dead time. 
Since the decay time of the FID is typically $\sim$30 $\mu$s, the time shifting of an FID placed at zero time may lead to a spurious FT spectrum.

Although a solid echo can be helpful for determining zero time of FID signals for dipolar-coupled systems~\cite{Slichter}, 
a recent analysis using a Gaussian back-extrapolation procedure for the obscured FID provides us with an alternative convenient method~\cite{GBNMR}.
The NMR spectrum from the composite FID is confirmed to be nearly identical to the magic-echo spectrum for some hydrides~\cite{GBNMR},
and then the extrapolation procedure has been applied to various hydrogen compounds~\cite{Corey1,Corey2,Corey3,Corey4}. 
 
In this paper, we adopted a replacement procedure to analyze the proton FID for Mg(OH)$_2$ and Ca(OH)$_2$,
which made the FT-NMR analysis better than that in a previous report~\cite{Itoh}.  
The portion of the FID obscured by probe ringing and/or dead time was replaced by the extrapolated data using an analytical fitting function. 
We report the detailed analysis of the FID and the broad components (intrinsic protons) in the NMR spectra for Mg(OH)$_2$ and Ca(OH)$_2$ powders,
and discuss possible relaxation mechanisms of the intrinsic proton spin-lattice relaxation rate 1/$T_1$. 

\section{FID Analysis} 
The observable part of the FID was extrapolated back to zero time using an analytical decay function, which jointed the extrapolated data to the part not obscured by probe ringing and/or dead time. Then, the resulting composite FIDs were analyzed by Fourier transformation.  
 
\begin{figure}[t]
 \begin{center} 
\includegraphics[width=1.05\linewidth]{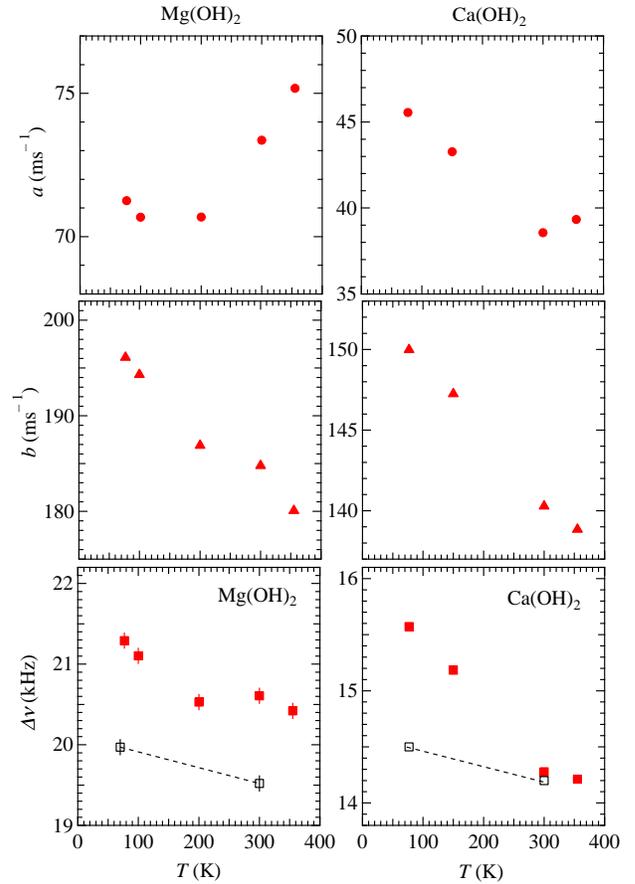}
 \end{center}
 \caption{\label{f2}
(Color online) Temperature dependences of decay rates of $a$, $b$, and a half-width $\Delta \nu$ for broad NMR spectra at $\nu_\mathrm{L}$ = 42.5772 MHz in Mg(OH)$_2$ (left) and Ca(OH)$_2$ (right). 
Open symbols are the theoretical values estimated from the rigid-lattice dipole fields through Van Vleck's second moment formula.  
The dashed lines are visual guides.
 }
\end{figure}       
 
We adopted the following modulated Gaussian function~\cite{Abragam},  
\begin{equation}
F(\tau)=F(0)\text{exp}\Bigl[-{a^2\tau^2\over {2}} \Bigr]{\text{sin} b\tau\over {b\tau}}, 
\label{eq1}
\end{equation}
where $F$(0), $a$, and $b$ are fitting parameters.  
The second moment $M_2$ of the frequency spectrum is given by
\begin{equation}
M_2=a^2 + {1 \over {3}}b^2,
\label{eq2}
\end{equation}
and the half-width of the frequency spectrum is given by
\begin{equation}
{\Delta \nu} = \sqrt{M_2}.  
\label{eq3}
\end{equation}     
 
In Fig.~\ref{f1}, the solid curves are the results from the least-squares fitting using an analytical function given by Eq.~(\ref{eq1}) for $\tau <$ 40 $\mu$s.
Figure~\ref{f2} shows the temperature dependences of the decay rates $a$ and $b$ for the short FID, and the half-width $\Delta \nu$ for NMR spectra in Mg(OH)$_2$ (left) and Ca(OH)$_2$ (right). 
Open symbols are the theoretical values estimated from Van Vleck's second moment formula for the homonuclear dipole coupling on a rigid lattice.
The lattice constants and the site parameters in the neutron diffraction data were adopted for the estimation~\cite{ND2,NDCa}.

The temperature dependence of $a$ in Mg(OH)$_2$ is different from that in Ca(OH)$_2$.
One should note that $a$ and $b$ are phenomenological parameters in the second moment of Eq.~(\ref{eq2})~\cite{Abragam}.  
Then, we have no theoretical accounts for the difference. 

The experimental half-widths $\Delta \nu$'s of Ca(OH)$_2$ are about 0.7 times those of Mg(OH)$_2$. 
This is reasonable, because the lattice constants of Ca(OH)$_2$ are longer than those of Mg(OH)$_2$~\cite{ND2,NDCa}. 
Qualitatively, the lattice expansion through Van Vleck's formula reproduces the decrease in $\Delta \nu$ from Mg(OH)$_2$ to Ca(OH)$_2$.
In Fig.~\ref{f2}, however, the experimental $\Delta \nu$ (closed symbols) of Ca(OH)$_2$ more steeply decreases with heating than the theoretical $\Delta \nu$ (open squares) on the rigid lattice.  

\begin{figure}[t]
 \begin{center}
 \includegraphics[width=0.90\linewidth]{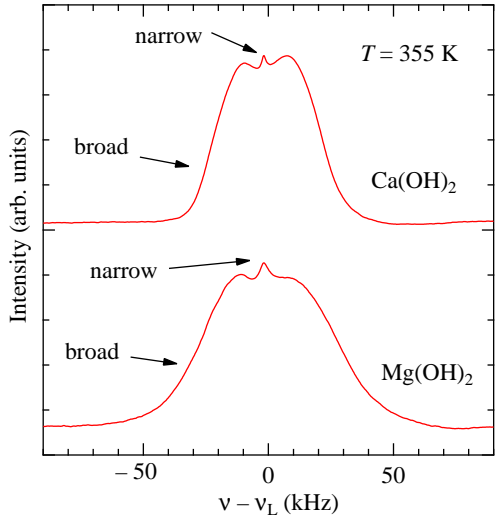}
 \end{center}
 \caption{\label{f4}
(Color online) FT-NMR spectra for Mg(OH)$_2$ and Ca(OH)$_2$ at 355 K and $\nu_\mathrm{L}$ = 42.5772 MHz. 
NMR spectra consist of broad and narrow components, which are assigned to intrinsic (immobile) and extrinsic (mobile) protons, respectively.    
The broad components are of rectangular type rather than Gaussian. 
 }
 \end{figure} 
 
The obscured parts of the short FIDs below 10 $\mu$s in Fig.~\ref{f1} were replaced by fitting Eq.~(\ref{eq1}) to the unobscured data. 
Figure~\ref{f4} shows the FT-NMR spectra of the extrapolated FIDs for Mg(OH)$_2$ and Ca(OH)$_2$ at 355 K and $\nu_\mathrm{L}$ = 42.5772 MHz. 
NMR spectra consist of broad and narrow components, which were assigned to intrinsic (immobile) and extrinsic (mobile) protons, respectively~\cite{Itoh}.    
The narrow spectra come from the long FIDs (60 $\mu$s $<$ $\tau <$ 2 ms)~\cite{Itoh2}, which are out of the frame of Fig.~\ref{f1}.   
The broad components of the pulsed FT-NMR spectra are compatible with the NMR absorption spectra measured by the continuous-wave method~\cite{cw1,cw2,cw3}.
Not simple Gaussian but rectangular-type line shapes characterize the broad components.
The intensities of the broad components relative to the narrow ones were underestimated in the original analysis~\cite{Itoh}
and were properly corrected in a previous report~\cite{Itoh2}.          
 
 % - - - - - - - - - - - - - - - - - - - - -
\section{Proton Spin-Lattice Relaxation Rates of Broad Components} 
 
 \begin{figure}[b]
 \begin{center}
 \includegraphics[width=0.95\linewidth]{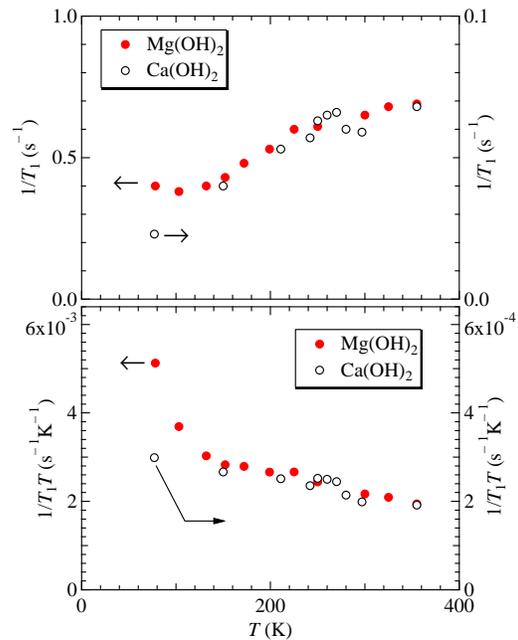}
 \end{center}
 \caption{\label{f5}
(Color online) Temperature dependences of the intrinsic proton spin-lattice relaxation rates 1/$T_1$'s (upper panel) and 1/$T_1T$'s (lower panel) at 42.5772 MHz for Mg(OH)$_2$ and Ca(OH)$_2$ powders, which are reproduced from previous reports~\cite{Itoh,Itoh2}.  
The magnitude of 1/$T_1$ for Ca(OH)$_2$ is about 0.1 times that for Mg(OH)$_2$.
 }
 \end{figure}       
 
 \begin{figure}[b]
 \begin{center}
 \includegraphics[width=0.85\linewidth]{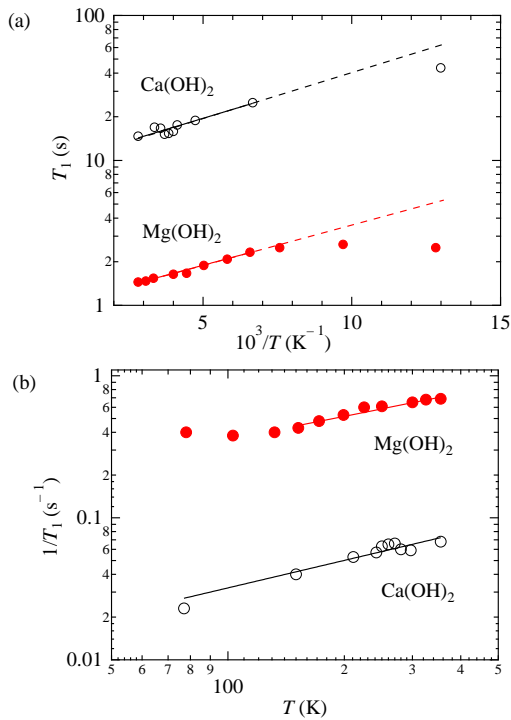}
 \end{center}
 \caption{\label{f6}
(Color online) (a) Semilog plots of the intrinsic proton spin-lattice relaxation times $T_1$'s against 10$^3$/$T$ for Mg(OH)$_2$ and Ca(OH)$_2$~\cite{Itoh,Itoh2}.
Solid lines are the results from the least-squares fitting using a thermal activation function of $T_1$ = $C$exp($E_g/k_\mathrm{B}T$): $E_g$ = 0.011 eV for Mg(OH)$_2$ and 0.012 eV for Ca(OH)$_2$. Dashed lines are visual guides.   
(b) Log$-$log plots of the intrinsic proton spin-lattice relaxation rates 1/$T_1$'s against temperature for Mg(OH)$_2$ and Ca(OH)$_2$~\cite{Itoh,Itoh2}.  
Solid lines are the results from the least-squares fitting using a power-law function of 1/$T_1$ = $AT^{n}$: $n$ = 0.54 for Mg(OH)$_2$ and 0.64 for Ca(OH)$_2$.
 }
 \end{figure} 
 
Figure~\ref{f5} shows the temperature dependences of the intrinsic proton spin-lattice relaxation rates 1/$T_1$'s (upper panel) and 1/$T_1T$'s (lower panel)  of the broad components for Mg(OH)$_2$ and Ca(OH)$_2$, which are reproduced from previous reports~\cite{Itoh,Itoh2}. 
The temperature dependence of 1/$T_1$ for Ca(OH)$_2$ is parallel to that below 200 K in Ref. 9. 

The concentration of a hypothetical impurity spin-1/2 was estimated to be at most 99 ppm for Mg(OH)$_2$ and 56 ppm for Ca(OH)$_2$
from Curie magnetism in the bulk magnetic susceptibility~\cite{SQUID}. 
The NMR measurement is a touchstone to test whether the impurity is substituted or mixed. 
Finite effects of the dilute magnetic impurities on the relaxation rates are excluded,  
because all the proton magnetizations in the broad component recover with a single exponential function for Mg(OH)$_2$~\cite{Itoh}.  
The stretched exponential relaxation with the variable exponent $\beta$ of $\sim$ 0.9 for Ca(OH)$_2$ may be due to some impurity effect~\cite{Itoh2}.
However, the fact that $T_1$ for Ca(OH)$_2$ is about 10 times longer than that for Mg(OH)$_2$ is in contrast to the impurity effect. 

The magnitude of 1/$T_1$ for Ca(OH)$_2$ is about 0.1 times that for Mg(OH)$_2$.
Both 1/$T_1T$'s monotonically decrease as temperature increases.
This is in contrast to the conventional Raman scattering of acoustic phonons, by which 1/$T_1T$ increases as temperature increases~\cite{Abragam}. 
  
In general, the proton spin-lattice relaxation rate 1/$T_1$ is proportional to the coupling constant of ($\Delta \nu$)$^2$ (second moment)~\cite{Abragam,BPP,Moriya}. 
$\Delta \nu$ for Ca(OH)$_2$ is about 0.7 times that for Mg(OH)$_2$.
Only the decrease of 0.49 in the coupling constant of ($\Delta \nu$)$^2$ is insufficient to account for the large decrease of 0.1 in 1/$T_1$.   
 
The higher atomic mass of the Ca ion than that of the Mg ion and the lattice expansion from Mg(OH)$_2$ to Ca(OH)$_2$ cause frequency shifts in lattice vibration modes, which in optical phonons were observed by Raman spectroscopy~\cite{Raman}.
The lattice expansion is also expected to lead to the lower Debye frequency of Ca(OH)$_2$ than that of Mg(OH)$_2$.
The Raman scattering of acoustic phonons leads to 1/$T_1$ $\propto$ $(\Delta \nu)^2\Theta_DT^{2}$ with a Debye cut-off frequency of $\Theta_D\sim a_\mathrm{H}^{-1}$ ($a_\mathrm{H}$ is the nearest-neighbor proton$-$proton distance)~\cite{Abragam}.
We estimated the decrease of $\sim$ 0.4 in $(\Delta \nu)^2\Theta_D$ from Mg(OH)$_2$ to Ca(OH)$_2$, 
which is insufficient to reproduce the decrease in 0.1 in 1/$T_1$.   

Figure~\ref{f6}(a) shows semilog plots of $T_1$'s against 10$^3$/$T$.
Solid lines are the results from the least-squares fitting using a thermal activation function of $T_1$ = $C$exp($E_g/k_\mathrm{B}T$) with the fitting parameters $C$ and $E_g$. 
We estimated the activation energies of $E_g$ = 0.011 eV for Mg(OH)$_2$ and 0.012 eV for Ca(OH)$_2$. 
These values are smaller than the previous NMR estimation of 0.21 eV for Ca(OH)$_2$~\cite{T1Ca}, 
the activation energy of 2.0 eV for the proton conductivity~\cite{Freund}, and 
the excitation energy of 0.4 eV for a proton jump in a local Morse potential~\cite{FreundB}. 
The energy $E_g$ may be the local hopping energy of the proton over three equivalent positions at the 6$i$ Wyckoff site ($x$, 2$x$, $z$) in $P\overline{3}$.

Figure~\ref{f6}(b) shows log$-$log plots of 1/$T_1$'s against temperature. 
The solid lines are the results from the least-squares fitting using a power-law function of 1/$T_1$ = $AT^{n}$ with the fitting parameters $A$ and $n$.
We estimated the powers of $n$ = 0.54 for Mg(OH)$_2$ and 0.64 for Ca(OH)$_2$, which are inconsistent with $n$ = 2 of the two-phonon Raman scattering of acoustic phonons~\cite{Abragam}. 
The powers of $n$ = 0.54 and 0.64 remind us of 1/$T_1\propto \sqrt{T}$ due to dilute electron gas in semiconductors~\cite{Abragam} and due to rattling phonons~\cite{Ueda2}. 

In Mg(OH)$_2$ and Ca(OH)$_2$, dilute mobile protons with Boltzmann statistics may act as the scattering carriers to the immobile protons. 
However, the $dc$ conductivity of Ca(OH)$_2$ is much higher than that of Mg(OH)$_2$~\cite{Freund}.
Since the carrier density $n_\mathrm{H}$ is in 1/$T_1\propto n_\mathrm{H}\sqrt{T}$~\cite{Abragam},
if Ca(OH)$_2$ involves a higher carrier density than Mg(OH)$_2$,  
the 1/$T_1$ of Ca(OH)$_2$ would be higher than that of Mg(OH)$_2$.
However, this is in contrast to the experimental results.  

Finally, we consider the effect of the anharmonic motion of the hydroxyl group due to broken site-symmetry~\cite{ND2,NDCa,DFT,MD}.
The neutron diffraction experiments indicate the large atomic displacement parameters of the protons~\cite{ND2,NDCa},
which are similar to those of the rattling ions in the pyrochlore~\cite{ZHiroi}. 
The anharmonic motion of the proton at the 6$i$ Wyckoff site ($x$, 2$x$, $z$) in $P\overline{3}$ can play a role in 1/$T_1\propto\sqrt{T}$ in the same way as the rattling phonons in the pyrochlore systems~\cite{Ueda2}.  

%------------------------
\section{Conclusions}

We analyzed the proton FIDs for Mg(OH)$_2$ and Ca(OH)$_2$ powders using an extrapolation function to the obscured parts of FIDs,
estimated the second moments of the intrinsic proton NMR spectra, and obtained the rectangular-type broad FT-NMR spectra.  
We found a large decrease in the magnitude of the intrinsic proton spin-lattice relaxation rate 1/$T_1$ from Mg(OH)$_2$ to Ca(OH)$_2$, which the decrease in the second moment is insufficient to account for.   
Local hopping and anharmonic rattling motions of the proton are the promising candidates for the proton spin-lattice relaxation mechanism.

%\begin{acknowledgment}
%\acknowledgment
%We thank M. Isobe for X-ray diffraction measurements and characterization of the powder samples. 
%\end{acknowledgment}


\begin{thebibliography}{9}
\bibitem{SQUID}A superconducting quantum interference device (SQUID) magnetometer was used to measure the bulk magnetic susceptibility for the present Mg(OH)$_2$ and Ca(OH)$_2$ powder in an external magnetic field of 1 T.
\end{thebibliography}
\end{document}